\documentclass[a4paper,twoside,10pt,reqno]{article}
\usepackage[english]{babel}
\usepackage[dvips]{graphicx}
\usepackage[T1]{fontenc}
\usepackage[latin1]{inputenc}
\usepackage{amssymb,amsmath,amsthm,dsfont,mathrsfs}
\usepackage{amsfonts,euscript}
\usepackage{color}
\usepackage{authblk}
\usepackage[dvipsnames]{xcolor}

\usepackage{mathtools}
\usepackage{tabls}
\usepackage{color, colortbl}

\usepackage{array}

\definecolor{LightCyan}{rgb}{0.88,1,1}

\newcommand{\nid}{\noindent }

\renewcommand{\abstract}[1]{{\small \noindent \textbf{Abstract:} #1\\}}

\begin{document}
\pagestyle{myheadings}


\title{A new blocks estimator for the extremal index}
\author{Helena Ferreira} 
\affil{Department of Mathematics of University of
	Beira
	Interior, Portugal\\ \texttt{helena.ferreira@ubi.pt}}

\author{Marta Ferreira} 
\affil{Center of Mathematics of Minho University\\ Center for Computational and Stochastic Mathematics of University of Lisbon \\
	Center of Statistics and Applications of University of Lisbon, Portugal\\ \texttt{msferreira@math.uminho.pt} }

\date{}

\maketitle

\abstract{The occurrence of successive extreme observations can have an impact on society. In extreme value theory there are parameters to evaluate the effect of clustering of high values, such as the extremal index. The estimation of the extremal index is a recurrent theme in the literature and there are several methodologies for this purpose. The majority of existing methods depend on two parameters whose choice affects the performance of the estimators. Here we consider a new estimator depending only on one of the parameters, thus contributing to a decrease in the degree of uncertainty. A simulation study presents motivating results. An application to financial data will also be presented.}

\nid\textbf{keywords:} {extreme value theory, stationary sequences, dependence conditions, extremal index}\\

\nid\textbf{AMS 2000 Subject Classification}  Primary: 60G70; Secondary: 62G32\\

\section{Introduction}\label{sintro}

Serial extremal dependence leads to the occurrence of clusters of high values. This is an  issue of major concern if associated to damaging phenomena, as for example, heatwaves whose duration in time can cause drought and wildfires. On the other hand, it may indicate a desirable situation, like  successive high stock returns attracting possible profits.

The extremal index, often denoted $\theta$, is a key parameter in assessing the extremal clustering degree. It ranges between $0$ and $1$, where smaller values mean stronger extremal dependence. Independent sequences have $\theta=1$ and no clustering of extremes takes place. A broad overview about the extremal index and, in particular, its applications in several areas can be seen in Moloney \emph{et al.} (\cite{moloney+19} 2019) and references therein.

One interpretation of $\theta$ is that it corresponds to the reciprocal limiting mean cluster size (Hsing et al.~\cite{hsing+88} 1988). The blocks and the runs estimators were developed upon this idea (Smith and Weissman \cite{smi+weis96} 1994, Weissman and Novak \cite{weis+98} 1998). 
Both estimators depend on the specification of two unknown parameters: a high threshold above which observations are considered extreme values and a cluster identifier. These are crucial since the methods show sensitiveness on their specification. 
In order to overcome the arbitrariness of these choices, alternative methods were proposed, such as the estimator of Ferro and Segers (\cite{ferro+segers03} 2003) which only needs the threshold specification or the estimator of Northrop (\cite{nor15} 2015) solely requiring the choice of the blocks length. 
Other estimation procedures can also be found in literature, such as the $K$-gaps estimator of S\"{u}veges and Davidson (2010) involving the choice of $K$ and of the threshold ($K=0$ leads to the estimator of Ferro and Segers \cite{ferro+segers03} 2003), estimators holding under a local dependence condition $D^{(s)}$ and thus requiring the indication of $s$ besides the threshold (Ferreira and Ferreira \cite{fer+fer18} 2018, Cai \cite{cai19} 2019), among others.

In this work we present a new estimator for the extremal index which only requires a block length parameter. Therefore, it intends to contribute to a decrease in the degree of uncertainty associated to the choice of parameters involving inference on $\theta$. 
The direct competitors are Ferro and Segers (\cite{ferro+segers03} 2003) and Northrop (\cite{nor15} 2015) estimators, although our methodology based on choosing a block size is closer to the second one. In Section \ref{sestim} we introduce the new proposal. Section \ref{ssim} addresses a simulation study in order to evaluate the performance of our estimator. In Section \ref{sapplic} we present an application to a financial time series. Final remarks and future work are pointed in Section \ref{sconc}.

\section{Methodology}\label{sestim}

Let $\{X_n\}_{n}$ be a stationary sequence with extremal index $\theta$ having, without loss of generality, standard Fr\'echet marginals, $F_{X}(x)=\exp(-1/x)$, $x>0$, and $\{\widehat{X}_n\}_{n}$ an associated iid sequence, i.e., an independent sequence having marginals also standard Fr\'echet, $F_{\widehat{X}}(x)=\exp(-1/x)$, $x>0$. Consider the bivariate sequence 
$$\{(Y_{n,1}=\widehat{X}_n,Y_{n,2}=(1/2)\widehat{X}_n\vee (1/2)X_n)\}_{n}\,.$$ 
We have that 
\begin{eqnarray}\nonumber
\begin{array}{rl}
&\displaystyle \lim_{n\to\infty} P\left(\bigvee_{i=1}^{n}Y_{i,1}\leq n/\tau_1,\bigvee_{i=1}^{n}Y_{i,2}\leq n/\tau_2\right)\\
=&\displaystyle \lim_{n\to\infty} P\left(\bigvee_{i=1}^{n}\widehat{X}_{i}\leq n/(\tau_1\vee(\tau_2/2))\right)P\left(\bigvee_{i=1}^{n}{X}_{i}\leq 2n/\tau_2\right)\vspace{0.25cm}\\
=& \exp\left(-\tau_1\vee(\tau_2/2)\right)\exp\left(-\theta\tau_2/2\right)\vspace{0.25cm}\\
=&\displaystyle (\exp(-\tau_1)\exp(-\theta\tau_2/2))\wedge \exp\left(-(1+\theta)\tau_2/2\right)\,.
\end{array}
\end{eqnarray}
Thus the limiting bivariate extreme value (BEV) copula is $C(u,v)=uv^{\frac{\theta}{1+\theta}}\wedge v$, which has tail dependence coefficient $\lambda_C$, given by
\begin{eqnarray}\label{lambda_theta}
\lambda_C=2-\displaystyle \lim_{u\to 1^{-1}}\frac{1-u^{1+\frac{\theta}{1+\theta}}\wedge u}{1-u} =1-\frac{\theta}{1+\theta}\,.
\end{eqnarray}
Our estimator is based on relation 
\begin{eqnarray}\label{theta_lambda}
\theta=\frac{1}{\lambda_C}-1
\end{eqnarray}
derived from (\ref{lambda_theta}). Since $\theta\in[0,1]$, then we must have $\lambda_C\in [1/2,1]$.

Estimators of the tail dependence coefficient of a random pair $(Z_1,Z_2)$ having a BEV df $G(x_1,x_2)=C_G(G_1(x_1),G_2(x_2))$ are addressed in literature with a threshold-free formulation. BEV copula $C_G$ can be stated as $C_G(G_1(x_1),G_2(x_2))=\exp(-l(-\log x_1,-\log x_2))$, where $l$ is the so called stable tail dependence function (Huang, \cite{huang92} 1992).

Consider $(Z_{1,1},Z_{1,2}),...,(Z_{n,1},Z_{n,2})$ a random sample of $(Z_1,Z_2)$ with BEV copula $C_G$ and stable tail dependence function $l$. We are going to use estimator
\begin{eqnarray}\label{lambda_stdf}
\widetilde{\lambda}=2-\widetilde{l}(1,1),
\end{eqnarray}
where 
\begin{eqnarray}\label{stdf}
	\widetilde{l}(1,1)=\frac{1}{1-\frac{1}{n}\sum_{i=1}^{n}\left(\widetilde{G}_1(Z_{i,1})\vee \widetilde{G}_2(Z_{i,2})\right)}-1
\end{eqnarray}
and $\widetilde{G}_j(x)=\frac{1}{n+1}\sum_{l=1}^n\mathds{1}_{\{Z_{l,j}\leq x\}}$, $j=1,2$, is the respective (modified) empirical df. 
Thus we have 
\begin{eqnarray}\label{est_lambda}
	\widetilde{\lambda}=3-\frac{1}{1-\frac{1}{n}\sum_{i=1}^{n}\left(\widetilde{G}_1(Z_{i,1})\vee \widetilde{G}_2(Z_{i,2})\right)},
\end{eqnarray}
and by (\ref{theta_lambda}) we obtain estimator
\begin{eqnarray}\label{est_theta}
\widetilde{\theta}=\frac{1}{\widetilde{\lambda}\vee 1/2}-1\,.
\end{eqnarray}
For more details on formulas (\ref{lambda_stdf}) and (\ref{stdf}) see Ferreira and Ferreira (\cite{fer+fer12a}, 2012a) and references therein. See also Ferreira and Ferreira (\cite{fer+fer12b} 2012b; \cite{fer+fer18} 2018).\\

The following algorithm describes our estimation proposal of the extremal index of a stationary sequence $X_1,...,X_n$. 
\begin{itemize}
	\item[Step 1.] In order to have standard Fréchet marginals, consider the marginal transformation $-\frac{1}{\log \widetilde{F}_{X}(X_i)}$, where $\widetilde{F}_{X}$ is the (modified) empirical df as defined above.
	\item[Step 2.] Generate an iid sequence with  standard Fréchet marginals, $\widehat{X}_1,...,\widehat{X}_n$, and consider random pairs, $(\widehat{X}_i,(1/2)\widehat{X}_i\vee (1/2)X_i)$, $i=1,...,n$.
	\item[Step 3.] Since we are going to first estimate $\lambda$ on the limiting BEV model of the component-wise maximum, we choose the blocks length $r$  where to take the component-wise maxima, in order to obtain a sample of maximums	
	$$\left(\bigvee_{i=(j-1)r+1}^{j\times r}\widehat{X}_i,\bigvee_{i=(j-1)r+1}^{j\times r}(1/2)\widehat{X}_i\vee (1/2)X_i\right),\,1\leq j\leq n/r.$$
	\item[Step 4.] Apply estimator $\widetilde{\lambda}$ given in (\ref{est_lambda}) on the random pairs of the previous step and calculate $\widetilde{\theta}$ in (\ref{est_theta}).
	\item[Step 5.] Repeat steps 2-5 a large number $M$ of times, obtain estimates $\widetilde{\theta}_{1},...,\widetilde{\theta}_{M}$ and take the mean, $\widetilde{\widetilde{\theta}}=\frac{1}{n}\sum_{s=1}^{M}\widetilde{\theta}_{s}$ in order to achieve robustness given the existence of arbitrariness in the generation of a random sample (step 2) in each estimate. Here we consider $M=10000$.
\end{itemize}
 
\section{Simulations}\label{ssim}
Our simulations are based on $100$ replicates of samples of size $1000$ and $5000$, of each of the following models: a first order autoregressive process with Cauchy marginals and autoregressive parameter $\rho=-0.6$ (Chernick \cite{chern78}, 1978), a negatively correlated uniform AR(1) process with $r=2$ (Chernick \emph{et al.} \cite{chern+91}, 1991), respectively denoted ARCau and ARUnif, a moving maxima (MM) process with coefficients $\alpha_0=2/6,\,\alpha_1=1/6,\,\alpha_2=3/6$ (Deheuvels \cite{deheuv83}, 1983), a first order MAR process with standard Fréchet marginals and autoregressive parameter $\phi=0.5$ (Davis and Resnick \cite{dav+res89}, 1989), a Markov chain (MC) with standard Gumbel marginals and logistic joint distribution with dependence parameter $\alpha=0.5$ (Smith \cite{smith92},  1992), an ARCH(1) process with Gaussian innovations, autoregressive parameter $\lambda=0.5$ and variance parameter $\beta=1.9\cdot10^{-5}$ (Embrechts \textit{et al.}, \cite{emb+97} 1997). The theoretical extremal index values of the processes ARCau, ARUnif, MM, MAR, MC and ARCH are, respectively, $0.64$, $0.75$, $0.5$, $0.5$, $0.328$ and $0.835$.

The root mean squared errors (rmse) and the absolute mean biases (abias) are given in Table \ref{tab1} for $n=1000$ and Table \ref{tab2} for $n=5000$.

For comparison, we consider two direct competitors of our estimator, as already mentioned in the Introduction, also requiring only one tuning parameter: the sliding blocks estimator of Northrop (\cite{nor15}, 2015) based on a block length choice and the estimator of Ferro and Segers (\cite{ferro+segers03}, 2003) which  depends on the choice of the high threshold. The first is denoted $\widetilde{\theta}^{N}$ and is computed for the same block lengths $r=10,20,30,40,50,70$ used in our proposal $\widetilde{\widetilde{\theta}}$. Ferro and Segers estimator is denoted $\widetilde{\theta}^{FS}$ and obtained for levels $u_n$ corresponding to the empirical quantiles $0.9$, $0.95$ and $0.99$, with respective notation, $q_{0.90}$, $q_{0.95}$ and $q_{0.99}$. 

Estimator $\widetilde{\widetilde{\theta}}$ seems to be competitive, particularly in models ARUnif, ARCau and ARCH.


\section{Application to financial data}\label{sapplic}

\begin{figure}[!h]
\begin{center}
\includegraphics[width=7.9cm,height=7.9cm]{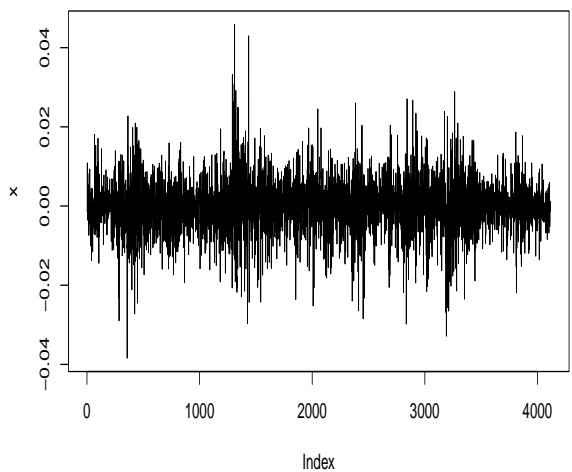}
\caption{Log-returns of the exchange rate US dollar versus UK pound, from January 2 of 1980 to May 21 of 1996.\label{fig1}}
\end{center}
\end{figure}

The data consists of the log-returns of the exchange rate US dollar versus UK pound, from January 2 of 1980 to May 21 of 1996 (Figure \ref{fig1}). An ARCH(1) fit was performed in Embrechts \textit{et al.}, \cite{emb+97} (1997) leading to the theoretical $\theta=0.835$.

\begin{figure}[!h]
	\begin{center}
		\includegraphics[width=7.9cm,height=7.9cm]{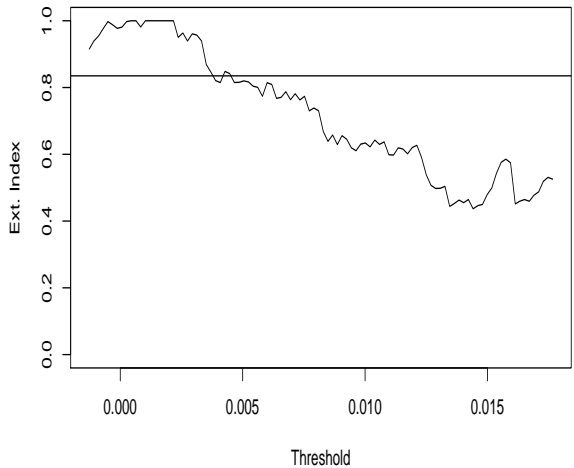}
		\caption{Plot of the $\widetilde{\theta}^{FS}$ estimates for thresholds  corresponding to sample quantiles ranging from $40\%$ to $99\%$. The horizontal line corresponds to theoretical $\theta=0.835$.\label{fig2}}
	\end{center}
\end{figure}

Figure \ref{fig2} presents estimates from $\widetilde{\theta}^{FS}$ computed for thresholds corresponding to sample  percentiles ranging from $40\%$ to $99\%$. The estimated values tend to decrease as the threshold increases falling below the theoretical one (horizontal line) from approximately quantile $86\%$.

Estimates obtained from $\widetilde{\widetilde{\theta}}$ and $\widetilde{\theta}^{N}$, for $r=10,20,30,40,50,70$, can be seen in Table \ref{tabData}. The results of $\widetilde{\widetilde{\theta}}$ are overall closer to the theoretical $\theta=0.835$.


\begin{table}[!h]
	\renewcommand\thetable{3}
	\caption{Extremal index estimates of the log-returns of the exchange rate US dollar versus UK pound, from January 2 of 1980 to May 21 of 1996, obtained for estimators $\widetilde{\widetilde{\theta}}$ and $\widetilde{\theta}^{N}$, with block lengths $r=10,20,30,40,50,70$.
		\label{tabData}}
	\begin{center}
		{\tablinesep=2ex\tabcolsep=10pt
			\setlength{\extrarowheight}{0.25cm}
			\begin{tabular}{|l|cccccc|}
				\hline
				\hline
				&  $r=10$ & $r=20$ & $r=30$ & $r=40$ & $r=50$ & $r=70$ \\\hline
				$\widetilde{\widetilde{\theta}}$ & 0.861 & 0.762 & 0.695 & 0.669 & 0.662 & 0.607\\
				$\widetilde{\theta}^{N}$ & 0.891 & 0.760 & 0.695 & 0.647 & 0.608 & 0.561\\
				\hline
				\hline
			\end{tabular}
		}	
	\end{center}
\end{table}

\section{Conclusion}\label{sconc}

The idea of relating the extremal index with the tail dependence coefficient is not new. For instance, in Ferreira and Ferreira (\cite{fer+fer12c}, 2012c), $\theta$ was derived as a linear combination of lag-$m$ serial tail dependence coefficients, under some local dependence conditions. 

The new proposed estimator of the extremal index is based on a relation between $\theta$ and the tail dependence coefficient $\lambda$ of a BEV copula, without any assumptions on the dependence between the variables of the sequence that has extremal index. 
This work shows that once we find a relation between the extremal index and the tail dependence coefficient of some BEV copula, we can always explore it to obtain other estimators of $\theta$ from $\lambda$ estimation. Depending on the bivariate sequences we use to explore that relation, we may find estimators that will work better for some class of models than others. This approach opens up new avenues of investigation for the estimation of the extremal index.

 \begin{table}
 	\renewcommand\thetable{1}
	\caption{The root mean squared error (rmse) and the absolute mean bias (abias) obtained for estimator $\widetilde{\widetilde{\theta}}$ (with block lengths $r=10,20,30,40,50,70$ and $n=1000$) and estimator $\widetilde{\theta}^{FS}$ (with levels $u_n$ corresponding to the empirical quantiles $0.9$, $0.95$ and $0.99$, respectively denoted, $q_{0.90}$, $q_{0.95}$ and $q_{0.99}$). The results in bold correspond to the best performance and the italic denotes the second best performance within each model.
		\label{tab1}}
	\begin{center}
		{\tablinesep=2ex\tabcolsep=10pt
			\setlength{\extrarowheight}{0.2cm}
			\begin{tabular}{|l|cccccc|}
				\hline
				\hline
				rmse &  MAR & MM & ARUnif & ARCau & ARCH & MC \\\hline
				$\widetilde{\widetilde{\theta}}$ ($r=10$)& 0.097 &  0.082 & 0.105 & 0.101 &  \textbf{0.058} & {0.140} \\
				$\widetilde{\widetilde{\theta}}$ ($r=20$)&  {0.093} & {0.069} & \textbf{0.068} & \textit{0.078} & 0.090 & 0.142 \\	
				$\widetilde{\widetilde{\theta}}$ ($r=30$)& 0.106 & 0.072 & \textit{0.071} & \textit{0.078} & 0.122 & 0.122 \\
				$\widetilde{\widetilde{\theta}}$ ($r=40$)& 0.107 & 0.066 & \textbf{0.068} & \textbf{0.075} & 0.145 & 0.159\\
				$\widetilde{\widetilde{\theta}}$ ($r=50$)& 0.112 & {0.072} & {0.091} & {0.080} & {0.161} & 0.164 \\
				$\widetilde{\widetilde{\theta}}$ ($r=70$)& 0.107 & 0.084 & 0.129 & 0.090 & 0.195 & 0.172\\
				\hline
				$\widetilde{\theta}^{N}$ ($r=10$)& \textit{0.060} & 0.069 & 0.250 & 0.195 & 0.083 & {0.108} \\
				$\widetilde{\theta}^{N}$ ($r=20$)& \textbf{0.054} & \textbf{0.048} & 0.219 & 0.111 & \textit{0.079} & \textbf{0.081} \\
				$\widetilde{\theta}^{N}$ ($r=30$)& {0.064} & \textit{0.051} & 0.186 & 0.099 & 0.092 & \textit{0.083} \\
				$\widetilde{\theta}^{N}$ ($r=40$)& 0.070 & 0.061 & 0.168 & 0.100 & 0.100 & 0.087 \\
				$\widetilde{\theta}^{N}$ ($r=50$)& 0.074 &   0.075 & 0.157 & 0.107 & 0.108 & 0.091 \\
				$\widetilde{\theta}^{N}$ ($r=70$)& 0.090 & 0.091 & 0.146 & 0.128 & 0.125 & 0.108 \\
				\hline
				$\widetilde{\theta}^{FS}$ ($q_{0.90}$) & {0.080} & {0.075} & 0.239 & 0.125 & 0.119 & {0.088}\\
				$\widetilde{\theta}^{FS}$ ($q_{0.95}$) & 0.100 & 0.111 & 0.201 & 0.151 &  0.122 & 0.165 \\
				$\widetilde{\theta}^{FS}$ ($q_{0.99}$) & 0.250 & 0.201 & 0.208& 0.231 & 0.165 & 0.323 \\
				
				\hline
				\hline
				abias & MAR & MM & ARUnif & ARCau & ARCH & MC \\
				\hline
				$\widetilde{\widetilde{\theta}}$ ($r=10$)& 0.066 & 0.070 & 0.097 & 0.085 & {0.024} & 0.119 \\
				$\widetilde{\widetilde{\theta}}$ ($r=20$)& 0.046 & 0.041 & {0.040} & {0.036} & 0.061 & {0.098} \\
				$\widetilde{\widetilde{\theta}}$ ($r=30$)& 0.041 & 0.033 & \textbf{0.006} & \textit{0.013} & 0.096 & 0.096 \\
				$\widetilde{\widetilde{\theta}}$ ($r=40$)& 0.037 & 0.029 & \textit{0.019} & \textbf{0.000} & 0.120 & {0.096} \\
				$\widetilde{\widetilde{\theta}}$ ($r=50$)& 0.033 & 0.025 & 0.049 & 0.009 & 0.140 & 0.103 \\
				$\widetilde{\widetilde{\theta}}$ ($r=70$)& 0.024 & 0.013 & 0.101 & 0.039 & 0.180 & 0.110 \\
				\hline
				$\widetilde{\theta}^{N}$ ($r=10$)& 0.052 & 0.062 & 0.250 & 0.191 & 0.068 & 0.102 \\
				$\widetilde{\theta}^{N}$ ($r=20$)& 0.027 & 0.024 & 0.215 & 0.094 & 0.032 & 0.063 \\
				$\widetilde{\theta}^{N}$ ($r=30$)& 0.019 & 0.012 & 0.173 & 0.064 & 0.015 & 0.051 \\
				$\widetilde{\theta}^{N}$ ($r=40$)& 0.013 & 0.008 & 0.143 & 0.048 & \textit{0.006} & 0.045 \\
				$\widetilde{\theta}^{N}$ ($r=50$)& \textbf{0.007} & \textit{0.006} & 0.122 & 0.040 & \textbf{0.003} & 0.040  \\
				$\widetilde{\theta}^{N}$ ($r=70$)& \textit{0.011} & \textbf{0.001} & 0.093 & 0.035 & 0.018 & \textit{0.039}  \\
				\hline
				$\widetilde{\theta}^{FS}$ ($q_{0.90}$) & {0.014} & {0.010} & 0.232 & 0.079 & 0.031 & \textbf{0.036} \\
				$\widetilde{\theta}^{FS}$ ($q_{0.95}$) & {0.019} & {0.015} & 0.178 & 0.056 & {0.024} & 0.068 \\
				$\widetilde{\theta}^{FS}$ ($q_{0.99}$) & 0.124 & 0.103 & 0.133 & 0.130 & {0.030} & 0.219 \\
				
				\hline
				\hline
			\end{tabular}
		}
	\end{center}
\end{table}

\begin{table}
	\renewcommand\thetable{2}
	\caption{The root mean squared error (rmse) and the absolute mean bias (abias) obtained for estimator $\widetilde{\widetilde{\theta}}$ (with block lengths $r=10,20,30,40,50,70$ and $n=5000$) and estimator $\widetilde{\theta}^{FS}$ (with levels $u_n$ corresponding to the empirical quantiles $0.9$, $0.95$ and $0.99$, respectively denoted, $q_{0.90}$, $q_{0.95}$ and $q_{0.99}$). The results in bold correspond to the best performance and the italic denotes the second best performance within each model.
		\label{tab2}}
	\begin{center}
		{\tablinesep=2ex\tabcolsep=10pt
			\setlength{\extrarowheight}{0.2cm}
			\begin{tabular}{|l|cccccc|}
				\hline
				\hline
				rmse &  MAR & MM & ARUnif & ARCau & ARCH & MC \\\hline
				$\widetilde{\widetilde{\theta}}$ ($r=10$)& {0.055} & 0.071 & 0.136 & 0.092 & \textbf{0.033} & {0.100}\\
				$\widetilde{\widetilde{\theta}}$ ($r=20$)& {0.053} & {0.045} & {0.088} & 0.064 & \textit{0.043} & {0.077}\\
				$\widetilde{\widetilde{\theta}}$ ($r=30$)& 0.061 & {0.038} & 0.072 & 0.062 & 0.053 & 0.077 \\
				$\widetilde{\widetilde{\theta}}$ ($r=40$)& 0.069 & {0.040} & \textit{0.061} & 0.067 & 0.066 &  0.081 \\
				$\widetilde{\widetilde{\theta}}$ ($r=50$)& 0.074 & {0.045} & \textbf{0.055} & 0.072 & 0.076 & 0.086 \\
				$\widetilde{\widetilde{\theta}}$ ($r=70$)& 0.089 & 0.051 & \textbf{0.055} & 0.072 & 0.090 & 0.099\\
				\hline
				$\widetilde{\theta}^{N}$ ($r=10$)& 0.050 & 0.068 & 0.250 & 0.188 & 0.074 & {0.105} \\
				$\widetilde{\theta}^{N}$ ($r=20$)& \textit{0.031} & {0.038} & 0.223 & 0.093 & {0.045} & {0.063} \\
				$\widetilde{\theta}^{N}$ ($r=30$)& \textbf{0.030} & \textit{0.030} & 0.177 & 0.066 & \textit{0.044} & 0.055 \\
				$\widetilde{\theta}^{N}$ ($r=40$)& {0.032} & \textbf{0.029} & 0.153 & \textit{0.054} & {0.046} & 0.045 \\
				$\widetilde{\theta}^{N}$ ($r=50$)& {0.034} & {0.031} & 0.136 & \textbf{0.050} & 0.050 & \textit{0.043} \\
				$\widetilde{\theta}^{N}$ ($r=70$)& {0.040} & {0.034} & 0.116 & \textit{0.054} & 0.059 & 0.045 \\
				\hline
				$\widetilde{\theta}^{FS}$ ($q_{0.90}$) &  {0.041} & {0.035} & 0.247 & 0.077 & 0.057 & \textbf{0.032}\\
				$\widetilde{\theta}^{FS}$ ($q_{0.95}$) & {0.047} & {0.049} & 0.170 & {0.057} & 0.065 & 0.045\\
				$\widetilde{\theta}^{FS}$ ($q_{0.99}$) & 0.119 & 0.083 & 0.145 & 0.131 & 0.115 & 0.118\\
				
				\hline
				\hline
				abias & MAR & MM & ARUnif & ARCau & ARCH & MC \\
				\hline
				$\widetilde{\widetilde{\theta}}$ ($r=10$)& 0.047 & 0.068 & 0.133 & 0.088 & 0.017 & {0.045} \\
				$\widetilde{\widetilde{\theta}}$ ($r=20$)& 0.027 & 0.035 & {0.078} & 0.046 & {0.008} &  {0.056}  \\
				$\widetilde{\widetilde{\theta}}$ ($r=30$) & 0.022 & 0.024 & 0.057 & 0.034 & 0.021 & 0.095 \\
				$\widetilde{\widetilde{\theta}}$ ($r=40$)& 0.023 & 0.020 & 0.042 & 0.031 & 0.037 &  0.041 \\
				$\widetilde{\widetilde{\theta}}$ ($r=50$)& 0.024 & 0.023 & \textit{0.027} & 0.027 & 0.045 & 0.040 \\
				$\widetilde{\widetilde{\theta}}$ ($r=70$)& 0.021 & 0.021 & \textbf{0.014} & \textbf{0.018} & 0.060 & 0.046 \\
				\hline
				$\widetilde{\theta}^{N}$ ($r=10$)& 0.048 & 0.067 & 0.250 & 0.187 & 0.071 & {0.018} \\
				$\widetilde{\theta}^{N}$ ($r=20$)& 0.024 & 0.034 & 0.222 & 0.088 & 0.033 & {0.024} \\
				$\widetilde{\theta}^{N}$ ($r=30$)& 0.015 & 0.021 & 0.174 & 0.055 & 0.018 & 0.040 \\
				$\widetilde{\theta}^{N}$ ($r=40$)& 0.010 & 0.016 & 0.148 & 0.038 & 0.010 & 0.049 \\
				$\widetilde{\theta}^{N}$ ($r=50$)& \textit{0.006} & 0.013 & 0.128 & 0.027 & \textbf{0.004} & 0.054 \\
				$\widetilde{\theta}^{N}$ ($r=70$)& \textbf{0.000} & 0.008 & 0.103 & \textbf{0.018} & \textit{0.005} & 0.063 \\
				\hline
				$\widetilde{\theta}^{FS}$ ($q_{0.90}$) & {0.015} & \textbf{0.000} & 0.245 & 0.057 & 0.024 & \textit{0.014}  \\
				$\widetilde{\theta}^{FS}$ ($q_{0.95}$) & {0.007} & \textit{0.005} & 0.156 & \textit{0.021} & 0.011 & \textbf{0.011} \\
				$\widetilde{\theta}^{FS}$ ($q_{0.99}$) & 0.044 & \textit{0.005} & {0.087} & 0.053 & 0.016 & 0.048 \\
				
				\hline
				\hline
			\end{tabular}
		}
	\end{center}
\end{table}


\begin{thebibliography}{000}
	
	
	\bibitem{cai19}Cai, J.J. (2019). A nonparametric estimator of the extremal index, arXiv 1911.06674
	
	\bibitem{chern78} Chernick M.R. (1978). Mixing conditions and limit theorems for maxima of some stationary sequences. PhD dissertation, Stanford University.
	
	\bibitem{chern+91} Chernick M.R., Hsing T., McCormick W.P. (1991). Calculating the extremal index for a class of stationary sequences. Advances in Applied Probability 23, 835--850.
	
	
	\bibitem{dav+res89}Davis R., Resnick S. (1989). Basic properties and prediction of max-ARMA processes. Advances in Applied Probability 21, 781--803.
	
	\bibitem{deheuv83} Deheuvels P. (1983). Point processes and multivariate extreme values. Journal of Multivariate Analysis 13, 257--272.
	
	\bibitem{emb+97} Embrechts, P., Kl\"{u}ppelberg, C., Mikosch, T. (1997). Modelling Extremal Events. Springer, Berlin.
	
	
	
	\bibitem{fer+fer12a} Ferreira, H., Ferreira, M. (2012a). On extremal dependence of block vectors. Kybernetika 48(5), 988--1006.
	
	\bibitem{fer+fer12b} Ferreira, H., Ferreira, M. (2012b). Fragility Index of block tailed vectors. J. Statist. Plann. Inference 142(7), 1837--1848.
	
	\bibitem{fer+fer12c} Ferreira M., Ferreira H. (2012c). On extremal dependence: some contributions. TEST 21(3), 566--583.
	
	\bibitem{fer+fer18} Ferreira, H., Ferreira, M. (2018). Estimating the extremal index through local dependence. Annales de l?Institut Henri Poincaré - Probabilités et Statistiques, 54(2), 587--605. 
	
	\bibitem{ferro+segers03} Ferro C.A.T., Segers J. (2003). Inference for clusters of extreme values. Journal of the Royal Statistical Society: Series B  65, 545--556.
	
	
	
	
	\bibitem{huang92} Huang, X. (1992). Statistics of bivariate extreme values. PhD thesis, Erasmus University Rotterdam, Tinbergen, Institute Research Series 22.
	
	
	\bibitem{hsing+88} Hsing T., H\"{u}sler J., Leadbetter M.R. (1988). On the exceedance point process for a stationary sequence. Probability Theory and Related Fields  78, 97--112.
	
	
	
	
	
	%
	%
	%
	%
	%
	%
	\bibitem{moloney+19} Moloney, N.R., Faranda, D., Sato, Y. (2019). An overview of the extremal index. Chaos 29, 022101.
	
	\bibitem{nor15}Northrop, P.J. (2015). An efficient semiparametric maxima estimator of the extremal index. Extremes 18(4), 585--603.
	%
	%
	%
	%
	%
	%
	\bibitem{smith92} Smith, R. (1992). The Extremal Index for a Markov Chain. Journal of Applied Probability 29(1), 37-45. 
	
	\bibitem{smi+weis96} Smith R.L., Weissman I. (1996). Characterization and estimation of the multivariate extremal index. Technical report, University of North Carolina at Chapel Hill, NC.
	%
	%
	%
	\bibitem{weis+98} Weissman I., Novak S.Y. (1998). On blocks and runs estimators of the extremal index. Journal of Statistical Planning and Inference 66(2), 281--288.
	%
	%
\end{thebibliography}
\end{document}